\documentclass[a4paper,11pt]{article}

\usepackage{jheppub} 

\usepackage[T1]{fontenc} 
\newcommand{\be}{\begin{equation}}\newcommand{\ee}{\end{equation}}%
\newcommand{\bd}{\begin{displaymath}}\newcommand{\ed}{\end{displaymath}}
\newcommand{\bit}{\begin{itemize}}                                        
 \newcommand{\eit}{\end{itemize}}                                         
\newcommand{\ben}{\begin{enumerate}}                                      
 \newcommand{\een}{\end{enumerate}}                                       
\newcommand{\baa}{\begin{array}{lll}}                                     
 \newcommand{\eaa}{\end{array}}                                           
\newcommand{\ba}{\begin{eqnarray}}                                        
 \newcommand{\ea}{\end{eqnarray}}                                         
\newcommand{\Ds}{\displaystyle}                                           
\newcommand{\gev}[1]{\relax\ifmmode{\text{GeV}^{#1}}                      
                     \else{GeV$^{#1}${ }}\fi}                             
\def\MSbar{\relax\ifmmode\overline                                        
            {\rm MS}\else{$\overline{\rm MS}${ }}\fi}                     
\def\as{\relax\ifmmode \alpha_s\else{$ \alpha_s${ }}\fi}                  
\def\abar{\relax\ifmmode{\bar{a}}\else{$\bar{a}${ }}\fi}                  
      
\title{\boldmath Cut moments and a generalization of DGLAP equations}

\author[a,b,1]{D. Kotlorz, \note{Corresponding author.}}
\author[b]{S. V. Mikhailov}

\affiliation[a]{Opole University of Technology, Division of Physics,\\
                45-370 Opole, Ozimska 75, Poland}
\affiliation[b]{Bogoliubov Laboratory of Theoretical Physics, JINR,\\
                141980 Dubna, Russia}

\emailAdd{dorota@theor.jinr.ru}
\emailAdd{mikhs@theor.jinr.ru}

\abstract{
We elaborate a cut (truncated) Mellin moments (CMM) approach
that is constructed to study 
deep inelastic scattering in lepton-hadron
collisions at the natural kinematic constraints.
We show that generalized CMM obtained by
multiple integrations of the original parton distribution  $f(x,\mu^2)$
as well as ones obtained by multiple differentiations of this $f(x,\mu^2)$ also satisfy
the DGLAP equations with the correspondingly transformed evolution kernel $P(z)$.
Appropriate classes of CMM for the available experimental kinematic range
are suggested and analyzed.
Similar relations can be obtained for the structure functions $F(x)$,
being the Mellin convolution $F= C \ast f$, where $C$ is the coefficient
function of the process.
}

\begin{document} 
\maketitle
\flushbottom

\section{Introduction}
Deep inelastic  scattering (DIS) of leptons on hadrons 
providing unique information about the structure of the hadrons 
remains nowadays one of the best studied reactions.
It  tests also the scale evolution of the parton densities 
(named also distribution functions), 
one of the most important predictions of perturbative QCD (pQCD). 
Dependence on the argument $x$ of the parton density $f(x,\mu^2)$
is formed at a hadron scale by nonperturbative forces, 
while its dependence on factorization/renormalization scale $\mu$
can be obtained within pQCD.
The evolution of  $f(x,\mu^2)$ with $\mu^2$ is governed
by the  well-known  Dokshitzer-Gribov-Lipatov-Altarelli-Parisi (DGLAP)
evolution equation \cite{b1,b2,b3,b4}, presented in the space of the Bjorken variable $x$
($x=Q^2/(2(Pq))$, where $q$ -- transferred momentum, $-q^2=Q^2> 0$,
$P$ -- hadron momentum, $P^2=m^2$).
 
Alternatively, one can study how to evolve with scale $\mu^2~(\mu^2 \sim Q^2)$  
the Mellin moments of the parton densities, 
$f(\mu^2,n)\equiv \int_{0}^{1}f(x,\mu^2)x^{n-1} dx$.
It seems that these moments provide a natural framework of QCD analysis as they
originate from the basic formalism of operator product expansion (OPE).
However, the moments $f(\mu^2,n)$ appear as the result of idealization.
We need to invent new ``real'' observables of DIS, 
named the ``cut (truncated) Mellin moments'' (CMM)
 with a goal to overcome kinematic constraints naturally appearing 
in  real DIS experiments. 
Namely, the small values of the variable $x$ cannot be reached in experiment
 at bounded below transferred momentum $Q^2\geq Q_0^2>0$ 
and at not very large $2(Pq)=2mE$ $\sim$ transferred energy in the laboratory frame. 
The CMM $f(x_0,\mu^2,n)$ are  generalized moments of the parton density
$f(x,\mu^2)$  with lower limit of integration
$x_0 > 0$, $f(x,\mu^2) \to f(x_0,\mu^2,n)\equiv \int_{x_0}^{1}f(x,\mu^2)x^{n-1} dx$.
In this way, $f(x_0,\mu^2,n)$ in contrast with the standard $f(0,\mu^2,n)\equiv f(\mu^2,n)$
takes into account the kinematic constraint.

The actual requirement to deal with the cut moments appeared, e.g., 
for JLAB experiment EG1b \cite{2008Deur} on 
the Bjorken integral measured in polarized nonsinglet DIS investigation of parton 
density $g_1^{p}-g_1^{n}$. 
The values of the standard moment $g_1(0,\mu^2,1)$  of the parton density $g_1$ there, 
obtained by extrapolation of $g_1$ to the origin, 
differ $3\div 4$ times from the observable $g_1(x_0,\mu^2,1)$ that is really measured in the
experiment ( see Table~1 in \cite{2008Deur}) in a restricted region of $x$.
Looks evident that the procedure of extrapolation certainly reduces the accuracy of the results.
Otherwise, one would obtain how $g_1(x_0,\mu^2,1)$ is evolved with $\mu^2$ in contrast with the 
case of the conserved standard moment $g_1(\mu^2,1) \sim \Gamma^{p-n}_1(\mu^2)$. 

The idea of ``truncated'' Mellin moments of the parton densities in QCD
analysis was introduced and developed in the late 1990's \cite{b5,b6,b7,b8}. 
The authors obtained the nondiagonal differential evolution
equations, in which the $n$th truncated moment couples to all higher ones.
Later on, diagonal integro-differential DGLAP-type evolution equations for
the single and double truncated moments of the parton densities were
derived in \cite{b10} and \cite{b11,b12}, respectively.
The main finding of the truncated Mellin moments approach (CMMA) is that the $n$th
moment of the parton density obeys the DGLAP equation again, 
but with a rescaled evolution kernel $P_{1}(z)=z^n P(z)$ \cite{b10}.
The CMMA has already been successfully applied, e.g., in spin physics to derive
a generalization of the Wandzura-Wilczek relation in terms of the truncated
moments and to obtain the evolution equation for the structure
function, e.g., $g_2$ \cite{b12,DK PEPAN}.
Truncation of the moments in the upper limit is less important in comparison to the low-$x$
limit because of the rapid decrease of the parton densities as
$x\rightarrow 1$; nevertheless, a comprehensive theoretical analysis requires
an equal treatment of both truncated limits.
The evolution equations for double cut moments and their application to
study the quark-hadron duality were also discussed in \cite{b13}.
A evolution equations for CMM are universal -- they are valid
in each order of the pQCD expansion
and also for the singlet parton
distributions \cite{DK PEPAN}.
A similar generalization can be obtained for the structure functions $F(x)$,
$F= C \ast f$, where $C$ is the coefficient function of the process, 
signum $\ast$ means the Mellin convolution.
Indeed, the coefficient functions
$C(t)$ rescale in the same way as the evolutional kernels:
$C_1(z)=z^n C(z)$.

In this paper, we present a novel generalization of CMM $f(x, \mu^2,n)$ and 
the corresponding DGLAP equations for the nonsinglet case. 
We elaborate an approach to investigate within
a unified frame a smoothed parton density 
obtained by multiple integration of the initial $f(x,\mu^2)$ as well as
a sharpened one obtained by a multiple differentiation of this $f(x,\mu^2)$.
These generalized solutions in terms of CMM (gCMM) provide a powerful 
tool to study DIS processes at real kinematic constraints. 
Moreover, this gives new classes of parton gCMM for DIS description, whose analytical 
properties are discussed in detail for some interesting partial cases.
The main aim of this paper is to present an extended theoretical frame for gCMM
and to investigate analytic properties for them.
We only mention possible applications of these gCMM for the analysis of  
experimental data, preparing a specific analysis for an oncoming paper.
This paper is organized as follows: In Sec.~\ref{sec:general}, we formulate
the theoretical scheme used in this work. 
The detailed analysis of a multiple integrated (smoothed) CMM is 
presented in Sec.~\ref{sec: partial-1} by consideration of a methodically important
partial case.
We show that in this case the same defined gCMM cover multiple differentiated (sharpen) CMM. 
Following this line of consideration, in Sec.~\ref{sec: partial-2} we propose a gCMM that is appropriate 
for analysis of DIS sum rules. 
Our conclusions are drawn in Sec.~\ref{sec:concl},
while important technical details are collected in Appendix,
where we also discuss a restriction of this generalization.

\section{General solution of the DGLAP equation}
\label{sec:general}
The goal of this section is to construct new general solutions for the nonsinglet DGLAP equation
\begin{eqnarray}\label{eq:dglap1}
\mu^2\frac{d}{d\mu^2}f(z,\mu^2)\equiv \dot{f}(z,\mu^2)
&=& P \ast f(z)\equiv \int_{0}^{1}P(y) f(x,\mu^2)~\delta(z-x\cdot y) dy\,dx\,,
\end{eqnarray}
which are inspired by physically motivated CMM. 
The method of consideration will be illustrated  
via a brief derivation of the original CMM \cite{b10} in subsec.~\ref{sec:general_A}. 

\subsection{Evolution equations for original cut Mellin moments }
\label{sec:general_A}
Let us apply the integral transformation $\int_{z_1}^{1}z^{n-1}(\ldots)~dz$ to both sides
of the DGLAP equation (\ref{eq:dglap1}) taking into account that this transformation can be
represented in the form of Mellin convolution (Mc).
Introducing the notation
for the RHS of (\ref{eq:dglap1}) $\phi(z)\equiv P\ast f(z)$ and
using the Mc notation for the RHS of the transformed equation one obtains 
(the argument $\mu^2$ is omitted here and below) 
\begin{subequations}
\label{eq:dglap-1}
\begin{eqnarray}
\label{eq:dglap-1a}
 f(z) \to f_1(z_1;n)&\stackrel{def}{=}&\int_{z_1}^{1}z^{n-1} f(z) dz,  \\
\dot{f}_1(z_1;n)&=& \int_{z_1}^{1}z^{n-1}\phi(z) dz = \left( \frac{z_1^{n}}{t^{n}}\ast \phi(z) \right)(z_1)=
\left( \frac{z_1^{n}}{t^{n}}\ast \left(P\ast f_1\right) \right)(z_1)\,. \label{eq:dglap-1b}
\end{eqnarray}
 \end{subequations}
Taking into account known properties of Mellin convolution, 
which are discussed in detail in Appendix~\ref{App-A} 
(see the chain of Eqs.(\ref{eq:A4}) and Eq.(\ref{eq:A5}) ),
and applying them to the RHS of Eq.(\ref{eq:dglap-1b})
one gets the evolution equation for CMM $f_1$
\begin{eqnarray} 
  \label{eq:dglap-1c}
 \dot{f}_1(z;n)&=& P_1\ast f_1 (z),
   \end{eqnarray}
where  $ P_1(y)= P(y)\cdot y^{n}$. 
The same relation is evidently derived for the structure function (SF)
$F= C \ast f$ under this integral transform, $F \to F_1=  C_1 \ast f_1$,
where $C_1(t)=t^{n}\cdot C(t)$ \cite{b10,b11,b12}.

\subsection{Evolution equations for generalized cut Mellin moments }
Here we present a generalization of the results shown in the previous
section, obtained for multiintegration of the original function.
Namely, if $f(x,\mu^2)$ is a solution
of the nonsinglet DGLAP equation with the kernel $P(y)$: 
then the $k$-integrated function $f_k(z,\mu^2;\{n\}_{k})$ 
(the argument $\mu^2$ and the index $k$ is omitted below for simplicity)
\begin{eqnarray} \label{def:def1}
f(z;\{n\}_{k}) &\equiv& f(z;n_1,n_2, \ldots, n_k ) \nonumber \\
                    &=& \int_{z}^{1}z_{k}^{n_k-1}~dz_k
                    \int_{z_k}^{1}z_{k-1}^{n_{k-1}-1}~dz_{k-1}\cdot
 \ldots \cdot \int_{z_2}^{1}z_{1}^{n_1-1}~f(z_1,\mu^2)\,dz_1~,
\end{eqnarray}
which is a generalization of CMM and is also the solution of
the DGLAP equation 
\begin{subequations}
 \label{eq:dglap2}
\begin{equation}\label{eq:dglap2a}
\dot{f}(z;\{n\}_{k}) = {\cal P}\ast f(z;\{n\}_{k})\equiv 
\int_{0}^{1}{\cal P}(y) f(x;\{n\}_{k})~\delta(z-x\cdot y) dy\,dx
\end{equation}
with the kernel
\begin{equation}\label{eq:dglap2b}
{\cal P}(y)= P(y)\cdot y^{n_1+n_2+\ldots +n_k}\,. 
\end{equation}
 \end{subequations}
\textit{Proof. One should apply $k$ times the transformation (\ref{eq:dglap-1}) (or apply Lemma in Appendix A.) 
to both sides of DGLAP Eq.(\ref{eq:dglap1}). At the same transformation of SF $F=C\ast f$
one obtains the new SF ${\cal F}$ and the new coefficient function ${\cal C}$,
}
\begin{equation}\label{eq:dglapF}
F,~C \to {\cal F}= {\cal C}\ast f(z;\{n\}_{k}),~~{\cal C}=C(t)\cdot t^{n_1+n_2+\ldots +n_k}.
\end{equation}
 The generalized evolution
equation similar to Eq.(\ref{eq:dglap2}) can be obtained also for ${\cal F}$.
For this purpose, let us recall that the corresponding original
equation reads \cite{FLK81},
\begin{subequations}
\label{eq:dglap5}
\begin{eqnarray}\label{}
  \dot{F}(z;\mu^2)  &=& \left(K \ast F\right)(z); \\
  && K= P + \beta(a_s)\left(\partial_{a_s}C\right)\ast C^{-1}\,,
    \end{eqnarray}
     \end{subequations}
where $K$ is the modified kernel, while
$\beta$ is the QCD $\beta$-function, $\Ds \mu^2\frac{d}{d\mu^2} a_s(\mu^2)=\beta(a_s)$. 
Again, applying the integration in Eq.(\ref{eq:dglap-1}) $k$ times to Eq.(\ref{eq:dglap5}) 
one arrives at the evolution equation for ${\cal F}$  
 \begin{eqnarray}\label{eq:dglap6}
  \dot{{\cal F}}(z;\{n\}_{k})  &=&{\cal K}\ast {\cal F}(z;\{n\}_{k})
\end{eqnarray} 
with the kernel ${\cal K}(y)=K(y)\cdot y^{n_1+n_2+\ldots +n_k}$.

The general solution (\ref{def:def1}) is the source of various new partial
solutions and also already known results,
e.g., from (\ref{eq:dglap2}) at $k=1$ follows the original equation from
\cite{b10}, namely,
\be \label{eq:cor1}
\dot{f}_{1}(z;n_1) = \int_{0}^{1}\left[P(y)y^{n_1}\right]  f_{1}(x;n_1)~\delta(z-x\cdot y) dx dy.
\ee 
If one puts $z=0$ in (\ref{eq:cor1}) it reduces to
$$\dot{f}_{1}(0;n_1) = \left(\int_{0}^{1}P(y)y^{n_1-1}dy\right) \cdot f_{1}(0;n_1)\equiv -\gamma(n_1)\cdot f_{1}(0;n_1),$$
which is the  renormalization group equation for the standard moments $f(\mu^2,n_1)\equiv f_{1}(0;n_1)$
with the corresponding nonsinglet anomalous dimension $\gamma(n_1)= - \int_{0}^{1} P(y)y^{n_1-1}~dy$.

A simple way to explore the properties of gCMM in
Eq.(\ref{def:def1}) is to admit a single weight for all
integrations in the RHS of (\ref{def:def1}), namely, $n_i= \alpha$. 
Hence, after the integration one obtains
\begin{subequations} 
 \label{def:def1a}
 \begin{eqnarray} \label{def:def1a_a}
 f(z;\{\alpha\}_{k})&=& \int_{z}^{1}\left[\frac{t^{\alpha}-z^{\alpha}}{\alpha}\right]^{k-1}
                    \frac{ f(t)}{\Gamma(k)}~dt ~, \\
    {\cal P}(y)&=& P(y)\cdot y^{\alpha k };~~ {\cal C}(t)= C(t)\cdot t^{\alpha k }~.  \label{def:def1a_b}              
 \end{eqnarray}
\end{subequations} 
Based on Eqs.(\ref{def:def1a}),(\ref{eq:dglap2}) different interesting gCMM 
can be constructed.
The results of  two simplest cases, $\alpha=1,~0$, will be considered below.

\section{\!\! \!\!\!Generalized CMM solution of the DGLAP equation at $n_i=\alpha=1$}
\label{sec: partial-1}
In this section we, consider the partial solution of Eq.(\ref{def:def1a}) with $n_1=n$ 
and all others $n_k=\alpha=1,~k>1$,
\begin{subequations}
 \label{def:def2}
 \ba 
z^{n}f(z) \to f(z;\{n,1\}_{k}) \equiv  f(z;n,1, \ldots, 1 )&=& \int_{z}^{1}~dz_k\int_{z_k}^{1}~dz_{k-1}\cdot
 \ldots  \int_{z_2}^{1}z_{1}^{n}~f(z_1)dz_1 \nonumber\\
   &=&\int_{z}^{1}\frac{\left(t-z\right)}{\Gamma(k)}^{(k-1)}~t^{n}f(t)dt ,\label{def:def2a}\\
\text{with the kernel}~~{\cal P}(y)&=& P(y)\cdot y^{n+k}~.  \label{def:def2b} \\
f(z;\{n,1\}_{k}) \to z^{n}f(z) = \left(-\frac{d}{dz}\right)^k f(z;\{n,1\}_{k}) && \text{-- inverse operator}
\ea
 \end{subequations}
In comparison to Eq.(\ref{def:def1a}), in gCMM (\ref{def:def2}) and below
we use for further convenience additional weight $t^n$ at $f(t)$.
 Our aim here is to analyze solution (\ref{def:def2}) and to extend the range of definition of 
 the integer parameter $k$ to any real value $\nu$, $k \to \nu$.
 We shall show that the generalization of (\ref{def:def2a},~\ref{def:def2b}) for $\nu \to -k$ naturally 
 leads to a solution with few times differentiated  initial parton density $f(z)$.
 This result will give us an important methodical lesson for further consideration.
 
 \subsection{The continuation of the gCMM solution in real $k \to \nu \geq 0$}
Let us consider the kernel $\Ds \frac{\left(t-z\right)}{\Gamma(k)}^{(k-1)}$
of the integrand in (\ref{def:def2a} ) that accumulates all the dependence on  the $k$. 
This kernel enables us to generalize $f(z;\{\rho,1\}_{k})$ to any real index $k \to \nu$, 
$f(z;\{\rho,1\}_{\nu})$ can be defined as
\begin{subequations}
\label{def:def3}
 \ba 
 \frac{\left(t-z\right)}{\Gamma(k)}^{(k-1)}&\to & \frac{\left(t-z\right)}{\Gamma(\nu)}^{(\nu-1)} \label{def:def3a}\\
f(z;\{\rho,1\}_{k}) &\to&   f(z;\{\rho,1\}_{\nu}) \stackrel{def}{=} \int_{z}^{1}\frac{\left(t-z\right)}{
\Gamma(\nu) }^{(\nu-1)}
~t^{\rho}f(t)dt \label{def:def3b}
\ea
 \end{subequations}
with the DGLAP kernel $${\cal P}_{\nu}(y)=P(y)\cdot y^{\rho+\nu}.$$ \\
Using definition (\ref{def:def3b}) $f(z;\{\rho,1\}_{\nu})$ can be analytically extended at the point $\nu=0$. 
To show this, let term $t^{\rho}f(t)= \varphi_\rho(t)$ for shortcut notation and put $\nu=\varepsilon \to 0$, then
\ba\label{eq:cor2}
&&f(z;\{\rho,1\}_{0})= \lim_{\varepsilon \to 0}\left[f(z;\{\rho,1\}_{\varepsilon})\right]= \nonumber \\
&&\lim_{\varepsilon \to 0}\bigg[\int_{z}^{1}\frac{\left(t-z\right)}{\Gamma(\varepsilon)}^{(\varepsilon-1)}~
\varphi_\rho(t)dt =
\frac{1}{\Gamma(\varepsilon)}\int_{z}^{1}\frac{\varphi_\rho(t)-\varphi_\rho(z)}{\left(t-z\right)^{1-\varepsilon}}~dt
+\frac{\varphi_\rho(x)}{\Gamma(\varepsilon)}\int_{z}^{1}\frac{dt}{\left(t-z\right)^{1-\varepsilon}}= \nonumber \\
&&\varepsilon \int_{z}^{1}\frac{\varphi_\rho(t)-\varphi_\rho(z)}{\left(t-z\right)}dt + \varphi_\rho(z)\left(1+O(\varepsilon)\right)\bigg]
\stackrel{\varepsilon \to~0}{=}~\varphi_\rho(z).
\ea
\subsection{The continuation of the gCMM solutions to $\nu < 0$}  
Successively integrating by part the RHS of (\ref{def:def3b}) the $f(z;\{\rho,1\}_{\nu})$ can be 
 analytically extended to negative $\nu $. 
E.g., integrating by part the def. (\ref{def:def3}) for  $\nu\geq 0$ one obtains
 \ba \label{eq:cor3}
f(z;\{\rho,1\}_{\nu}) &=& \frac{\bar{z}^{\nu}}{\Gamma(\nu+1)}\varphi_\rho(1)-
\int_{z}^{1}~\varphi_\rho'(t)\frac{\left(t-z\right)^{\nu}}{\Gamma(\nu+1)}dt\, .
\ea
The RHS of Eq.(\ref{eq:cor3}) at $\nu=0$, 
\ba
f(z;\{\rho,1\}_{0})&=&\varphi_\rho(1)- \int_{z}^{1}~\varphi_\rho'(t)dt =\varphi_\rho(z)\, , 
\ea
coincides with the RHS of (\ref{eq:cor2}). 
By means of further integrations Eq. (\ref{eq:cor3}) can be extended to the left for $\nu > -1$.
Then, each of the integrations by part shifts to the left on $1$ domain of analyticity in $\nu $; 
finally, one arrives at the proposition:

\textit{
$f(z;\{\rho,1\}_{\nu})$ can be extended  into strip $\nu > -K-1$ at  any integer $K \geq0$ in the form}
\begin{subequations}
 \label{eq:dglap3}
  \begin{eqnarray} 
\!\!\!\!\!\! f(z;\{\rho,1\}_{\nu})  &\!\!=\!\!&  \sum_{m=1}^{K+1}\frac{\bar{z}^{(\nu+m-1)}}{\Gamma(\nu+m)}
  (-)^{m-1}\varphi_{\rho}^{(m-1)}(1) - 
   (-)^{K}\int_{z}^{1} \frac{\left(t-z\right)^{(\nu+K)}}{\Gamma(\nu+K)}\varphi_{\rho}^{(K+1)}(t) 
 dt,~~\label{eq:dglap3a}  \\
{\cal P}_{\nu}(y)&=&P(y)\cdot y^{\rho+\nu}~~\text{at}~~\rho+\nu  \geq 0;~~ 
\varphi_{\rho}^{(m)}(x)\equiv \frac{d^m}{dx^m}\varphi_{\rho}(x)
.\label{eq:kernel3}
\end{eqnarray}
 \end{subequations}
From Eq.(\ref{eq:dglap3a}) follows the expression for $f(z;\{\rho,1\}_{-K})$ at integer negative $\nu =-K$ 
(within the strip of extension $\nu > -K-1$). 
Put $\nu =-K+\varepsilon$  and taking the limit $\varepsilon \to 0$ in the RHS of (\ref{eq:dglap3a}) one arrives at
 \begin{eqnarray}
  \label{eq:neginteger1}
\lim_{\varepsilon \to 0} f(z;\{\rho,1\}_{-K+\varepsilon})= f(z;\{\rho,1\}_{-K})=(-)^{K} \varphi_{\rho}^{K}(z) .
\end{eqnarray} 
A few partial results for $f(x;\{\rho,1\}_{\nu})$ are shown in the table below
\begin{center}
\begin{tabular}{|r|c|c|c|c|c|} \hline 
$\nu=$&$2$         & $1$       & $0$      & $-1$      & $-2$    \\   \hline                                                
&                &               &              &               &     \\                  
$f(x;\{\rho,1\}_{\nu})=$&$\int_{x}^{1}dz\int_{z}^{1}\varphi_\rho^{}(t)~dt$                & $\int_{x}^{1}\varphi_{\rho}(t)dt$              
                            &$\varphi_\rho(x)$&$-\varphi_{\rho}'(x)$& $\varphi_{\rho}''(x)$  \\                   
&                &               &              &                &                                  \\          
${\cal P}_{\nu}=$&$P(y) y^{\rho+2}$&$P(y) y^{\rho+1}$&$P(y)y^{\rho}$        &$P(y) y^{\rho-1}$ &$P(y) y^{\rho-2}$ \\ \hline           

\end{tabular}                                                     
\end{center}                                                                       
An interesting case of solution provides the condition  \textit{$\rho~=\nu$~(number of derivatives)}.
 For any  $\nu\geq 0$,  $f(x;\{\nu,1\}_{-\nu})$ 
 from equation (\ref{eq:dglap3}) 
 evolves following to the DGLAP equation (\ref{eq:dglap1}) with the same initial kernel ${\cal P}_{\nu}= P$ 
 (see (\ref{eq:kernel3})) that has no $\nu$ dependence. 
The solutions  $\varphi_1^{(k)}(x)=\left(xf(x)\right)^{(k)}$, 
 were considered in \cite{OT-Winter05,A-T2009}.

We conclude that the  partial solution 
joints  different results, previously presented in the literature
 \cite{b10,b11,b12,OT-Winter05,A-T2009},
 in a unit frame: 
ones are related with the ``truncated integration'' of $f(x)$ in
\cite{b10,b11,b12}, 
while the other is related with differentiated  $f(x)$ in
 \cite{OT-Winter05,A-T2009}.
  The applicable domain of $f(x;\{\rho,1\}_{\nu})$ in $\nu$ 
 can be extended to the real axis, following  Eq.(\ref{eq:dglap3a}). 

\section{\!\!\!\!\! Generalized CMM solution of the DGLAP equation at $n_i=\alpha =0$} 
\label{sec: partial-2}
Here we construct a gCMM that would be appropriate for analysis of the DIS sum rules.
To this end, it is convenient to take  $f(z;\{n\}_{k})$ with a single DGLAP kernel ${\cal P}$ that 
is independent of $k$.
Following the condition Eq.(\ref{def:def1a_b}), let us consider the case of Eq.(\ref{def:def1a}) with $n_1=n$,
and all $n_k=\alpha=0,~k>1$,  then 
 $\Ds \lim_{\alpha \to 0}\left[\frac{t^{\alpha}-z^{\alpha}}{\alpha}\right]^{k-1}$ $= \ln^{(k-1)}\left(t/z\right)$,
\begin{subequations} 
 \ba 
 z^{n}f(z) \to f(z;\{n,0\}_{k}) \equiv f(z; \underbrace{n,0, \ldots, 0}_{k} )\!&\!=\!&\! 
\int_{z}^{1}\frac{\ln^{(k-1)}\left(t/z\right)}{\Gamma(k)}~t^{\rho}f(t)\frac{dt}{t} \label{def:def4},\\
\text{with the kernel}~{\cal P}(y)\!&\!=\!&\! P(y)y^{n},~\label{def:kern4}
\ea 
 \end{subequations}
This ~$y^{n}$-factor in the kernel appears due to the wight $z^{n}$ in $\varphi_n(x)=x^n f(x)$. \\
Moreover, the gCMM $f(z;\{1,0\}_{k})$ holds the same value of the normalization 
$\Ds \langle f\rangle = \int^1_0 f(x)dx$ as one has
for the parton density $f(x)$.
Indeed, gCMM can be represented as 
\ba
f(z;\{n,0\}_{k})&=& \int_{z}^{1}z_{k}^{-1}~dz_k\int_{z_k}^{1}z_{k-1}^{-1}~dz_{k-1}\cdot
 \ldots \cdot \int_{z_2}^{1}\varphi_n~z_{1}^{-1}dz_1 \\
 &\equiv&  \left(1\ast 1\ast \ldots \ast \varphi_n \right)(z),
\ea
therefore, $\langle f(z;\{n, 0\}_{k})\rangle =\prod^{k}_1  1\cdot \langle \varphi_n \rangle\mid_{n=1}
 =\langle f \rangle$. \\

Following the  previous proposition in Eq.(\ref{def:def3}),
we extend the solution (\ref{def:def4}) to any real $\nu > 0$ (and $\rho $)
\begin{subequations}
 \ba \label{def:def5}
f(z;\{\rho,0\}_{\nu}) &\stackrel{def}{=}& \int_{z}^{1}\frac{\ln^{(\nu-1)}\left(t/z\right)}{
\Gamma(\nu)}~~\varphi_{\rho}(t)
\frac{dt}{t}\, , \\
\text{with DGLAP kernel}&&{\cal P}(y)=P(y)\cdot y^{\rho}. \label{eq:ker5}
\ea
 \end{subequations}
The contribution to $f(z;\{\rho,0\}_{\nu})$ is  reinforced at the right end $t=1$  
in Eq.(\ref{def:def5}) by powers of logs. 
This reinforcement becomes especially useful for the case when the experimental data are better
known at larger $x$ and, in contrast, ones are unreliable or worse known at lower $x$.

An important property of gCMM $(\ref{def:def5})$ is the independence of
the corresponding kernel (\ref{eq:ker5}) of the parameter $\nu$. 
Therefore, 
collecting these solutions $f(z;\{\rho,0\}_{\nu})$ with the different wights we can obtain the new 
solution at the same kernel. 
In other words, the integrands $\ln^{(\nu-1)}\left(t/z\right)/\Gamma(\nu)$ at different $\nu$
can be considered as ``bricks'' for any new gCMM constructions that evolve following the same DGLAP
equation.

The next step of extension can be done like one in Eq.(\ref{eq:dglap3}) in the previous section:
$f(z;\{\rho,0\}_{\nu})$ can be extended  into strip $\nu > -K-1$ at  any integer $K$ in the form \\
\begin{subequations}
 \begin{eqnarray}\label{eq:dglap4}
\!\!\!\!\!\!  f(z;\{\rho,0\}_{\nu})  &\!\!=\!\!& \sum_{m=1}^{K+1}\frac{l^{(\nu+m-1)}}{\Gamma(\nu+m)}
  (-)^{(m-1)}\varphi_{\rho}^{(m-1)}[l] - 
   (-)^{K}\int_{0}^{l} \frac{y^{(\nu+K)}}{\Gamma(\nu+K+1)}\varphi_{\rho}^{(K+1)}[y] 
 dy, ~~~ \\
&&{\cal P}(t)=P(t)\cdot t^{\rho}~~\text{at}~~\rho\geq 0, \label{eq:kernel4}
\end{eqnarray}
 \end{subequations}
where $l=\ln(1/z)$ and $ y=\ln(t/z)$ is a new appropriate variable. 
It is convenient to invent the new notation for $\varphi_{\rho}$, 
$~\varphi_{\rho}[y]\equiv \varphi_{\rho}(ze^y)$ that depends on the variable $y$.
It is instructive to obtain from (\ref{eq:dglap4}) a partial case at $\nu \to -1$ (for the strip at $K=1$)
$$\lim_{\nu \to -1}f(z;\{\rho, 0\}_{\nu}) = f(z;\{\rho, 0\}_{-1}) = -z\frac{d}{dz} \varphi_{\rho}(z)\, .$$ 
In general, for any integer  $\Ds n>0: -n >  -K-1 $ one can obtain
\be
 \label{eq:neginteger0}
f(z;\{\rho,0\}_{-n}) = \Ds \left(-z \frac{d}{dz}\right)^{n}\varphi_{\rho}(z)\,,
\ee
the result is similar to one in Eq.(\ref{eq:neginteger1})
and also to that one obtained in \cite{OT-Winter05,A-T2009}.
While for any fractional $\Ds \nu>0: -\nu > -K-1$, $f(x;\{\rho,0\}_{-\nu})$
from equation (\ref{eq:dglap4}) can be considered as generalized derivatives 
$\Ds \left(-x \frac{d}{dx}\right)^{\nu}\varphi_{\rho}(x)$.

\section{Conclusion}  
 \label{sec:concl} 
We propose a generalization of the standard parton density $f(x)$ 
by means of the extended cut Mellin moments.
These CMM  are appropriate for direct study of real observables in DIS
at the natural kinematic constraints of experiments.
We found that functions obtained by multiple integration of the original
parton density $f(x,\mu^2)$ as well as ones obtained by multiple
differentiations of this $f(x,\mu^2)$ satisfy the same corresponding
DGLAP equations. 
Moreover, this gives new classes of parton generalized CMM for DIS descriptions 
where both the previous cases are considered  within a unified frame.
Some interesting partial cases can be especially useful when the experimental
data are better known at large $x$ and are less known at low $x$.
We showed also similar relations
for the structure functions $F(x)$, being the Mellin convolution $F= C \ast f$,
where $C$ is the coefficient function of the process. 
In other words, the presented here CMM approach provides novel gCMM functions,
satisfying the same DGLAP equations, which reinforce the experimentally
available $x$-region.
This seems to be of great importance in QCD analysis of a wide class of
high-energy processes. \\
Our goal here was to present an extended theoretical frame for gCMM.
We only mentioned possible applications of these gCMM while the
specific analysis of the experimental data is reserved for forthcoming papers.

\acknowledgments
We are grateful to O. Teryaev for the fruitful discussions of the considered subject 
which inspired this work.
DK would like to thank BLTP community for their warm hospitality and cozy
atmosphere.
This work is supported by the Bogoliubov-Infeld Program, Grant No
01-3-1113-2014/2018.
SVM acknowledges support from  the Russian Foundation for Fundamental
Research (Grant No.\ 14-01-00647a)
\begin{appendix}
\appendix
\section{Lemma about Mellin convolution}
\label{App-A}\setcounter{equation}{0}
Let us consider the RHS of the DGLAP equation $\phi(z)$
\begin{equation}
\label{eq:A1}
\mu^2\frac{d}{d\mu^2}f(z,\mu^2)=\phi(z) \equiv \left(P\ast f\right)(z)~,
\end{equation}
and define the map for any function $\phi$, $\phi \rightarrow \phi_1$ by means of the Mellin convolution
 \begin{eqnarray}
  \label{eq:A1a}
 \phi_1(z_1)&=&\int_{z_1}^{1}~\omega(z)\phi(z)dz =\Big( \omega\left( \frac{z_1}{t} \right) \frac{z_1}{t}\ast \phi \Big)(z_1)~.
  \end{eqnarray}
For the partial case of the monomial smearing function $\omega(z)=z^{n-1}$ one can write
\begin{subequations}
 \label{eq:A2}
  \begin{eqnarray}
\phi_1(z_1)&=&\int_{z_1}^{1}~z^{n-1}\phi(z)dz =\left( \frac{z_1^{n}}{t^{n}}\ast \phi \right)(z_1)
 \label{eq:A2a}, 
  \end{eqnarray}
  and for $k$--times recursion, used in (\ref{eq:dglap2}),
\begin{eqnarray}  
 \phi_{k+1}(z_{k+1})&=&\int_{z_{k+1}}^{1}~z^{n_k-1}\phi_{k}(z)dz~.
  \end{eqnarray}  
   \end{subequations}
What is the RHS of Eq.(\ref{eq:A2a})? Substituting (\ref{eq:A1}) in the RHS of (\ref{eq:A2a}) 
and using the commutative and associative properties 
of Mellin convolution one has
 \begin{eqnarray}
  \label{eq:A4}
 \phi_1(z_1)&=& \left(\frac{z_1^{n}}{t^{n}}\ast \left(P\ast f\right)\right)(z_1)=
  \left(\left(P\ast f\right)\ast\frac{z_1^{n}}{t^{n}} \right)(z_1)=
  \left(P\ast \left(f\ast\frac{z_1^{n}}{t^{n}}\right) \right)(z_1)= \nonumber \\
 &&\left(z_1^{n}P\ast \frac{1}{z^{n}}\left(f\ast\frac{z^{n}}{t^{n}}\right)(z) \right)(z_1)=
 \left(z_1^{n}P\ast \frac{1}{z^{n}}f_1(z;n) \right)(z_1)
 \end{eqnarray} 
  The variable $z$, explicitly invented in the intrinsic convolution 
  in the LHS of the last equation, is the argument of this intrinsic convolution,
 i.e., $\left(f\ast 1/t^{n}\right)(z)$. 
 In the RHS of (\ref{eq:A4}) there appears CMM $f_1(z_1;n)=\int_{z_1}^{1}z^{n-1}f(z)dz$ following Eq.(\ref{eq:A1}).
   As the last step one can obtain the transformed kernel $P_1$ in the convolution
 \begin{subequations}
  \label{eq:A5}
  \begin{eqnarray} 
  \label{eq:A5a}
  \phi_1(z_1)&=& \left(z_1^{n}P\ast \frac{1}{z^{n}}f_1(z;n) \right)(z_1)=\left(P_{1(n)}\ast f_1 \right)(z_1),\\
  && f_1(z_1,n)=\int_{z_1}^{1}z^{n-1}f(z)dz;~ P_{1(n)}(t)= P(t)t^{n},  \label{eq:A5a}
   \end{eqnarray} 
   \end{subequations}
where  the last equation for $P_1$ followed from the definition of the Mellin convolution in 
(\ref{eq:dglap1}). \\
 If $P$ is in turn $P= A\ast B$, then 
\begin{eqnarray} 
  \label{eq:A6}
 P_1(t)= \left(A_1\ast B_1\right)(t), 
\text{where}~ A_1(\alpha)=\alpha^n A(\alpha) ,~ B_1(\beta)=\beta^n B(\beta).
 \end{eqnarray}

Let us emphasize here that the chain of conclusions from Eq(\ref{eq:A1}), Eq(\ref{eq:A2a}) 
to Eq(\ref{eq:A5}) is possible only for monomial weight $\omega(z)=z^{n-1}$ 
in the definition of transform  in Eq(\ref{eq:A1a}).
 For another form of the weight $\omega$ we do not obtain a covariant form of the RHS of Eq.(\ref{eq:A1}).
 Really, substituting  the linear combination of the monomials in $w= a z^{n-1}+b z^{m-1}$ in 
 Eq.(\ref{eq:A1a}) one can obtain $\varphi_1(z_1)$ for the LHS of Eq.(\ref{eq:A1}),
 while for the RHS  one will arrive at the decomposition
 \ba
       a \left(P_{1(n)}\ast f_{1}(z;n)\right)(z_1)+ b \left(P_{1(m)}\ast f_{1}(z;m)\right)(z_1)
 \ea
that can not be presented as a single convolution with $\varphi_1$.
 
\end{appendix} 

\end{document}